\documentclass[]{aastex631}

\received{January 17, 2022}
\revised{March 29, 2022}
\accepted{March 31, 2022}
\shorttitle{Dust Ejection From (3200) Phaethon at Perihelion}
\shortauthors{Kimura et al.}

\begin{document}

\title{Electrostatic Dust Ejection From Asteroid (3200) Phaethon With the Aid of Mobile Alkali Ions at Perihelion}

\correspondingauthor{Hiroshi Kimura}
\email{hiroshi\_kimura@perc.it-chiba.ac.jp}

\author{Hiroshi Kimura}
\affiliation{Planetary Exploration Research Center (PERC), Chiba Institute of Technology, Tsudanuma 2-17-1, Narashino, Chiba 275-0016, Japan}

\author{Katsuhito Ohtsuka}
\affiliation{Tokyo Meteor Network, 1-27-5 Daisawa, Setagaya-ku, Tokyo, 155-0032, Japan}

\author{Shota Kikuchi}
\affiliation{Planetary Exploration Research Center (PERC), Chiba Institute of Technology, Tsudanuma 2-17-1, Narashino, Chiba 275-0016, Japan}

\author{Keiji Ohtsuki}
\affiliation{Graduate School of Science, Kobe University, 1-1 Rokkodai-cho, Nada-ku, Kobe 657-8501, Japan}

\author{Tomoko Arai}
\affiliation{Planetary Exploration Research Center (PERC), Chiba Institute of Technology, Tsudanuma 2-17-1, Narashino, Chiba 275-0016, Japan}

\author{Fumi Yoshida}
\affiliation{School of Medicine, Department of Basic Sciences, University of Occupational and Environmental Health, Japan, \\
1-1 Iseigaoka, Yahata, Kitakyusyu 807-8555, Japan}
\affiliation{Planetary Exploration Research Center (PERC), Chiba Institute of Technology, Tsudanuma 2-17-1, Narashino, Chiba 275-0016, Japan}

\author{Naoyuki Hirata}
\affiliation{Graduate School of Science, Kobe University, 1-1 Rokkodai-cho, Nada-ku, Kobe 657-8501, Japan}

\author{Hiroki Senshu}
\affiliation{Planetary Exploration Research Center (PERC), Chiba Institute of Technology, Tsudanuma 2-17-1, Narashino, Chiba 275-0016, Japan}

\author{Koji Wada}
\affiliation{Planetary Exploration Research Center (PERC), Chiba Institute of Technology, Tsudanuma 2-17-1, Narashino, Chiba 275-0016, Japan}

\author{Takayuki Hirai}
\affiliation{Planetary Exploration Research Center (PERC), Chiba Institute of Technology, Tsudanuma 2-17-1, Narashino, Chiba 275-0016, Japan}

\author{Peng K. Hong}
\affiliation{Planetary Exploration Research Center (PERC), Chiba Institute of Technology, Tsudanuma 2-17-1, Narashino, Chiba 275-0016, Japan}

\author{Masanori Kobayashi}
\affiliation{Planetary Exploration Research Center (PERC), Chiba Institute of Technology, Tsudanuma 2-17-1, Narashino, Chiba 275-0016, Japan}

\author{Ko Ishibashi}
\affiliation{Planetary Exploration Research Center (PERC), Chiba Institute of Technology, Tsudanuma 2-17-1, Narashino, Chiba 275-0016, Japan}

\author{Manabu Yamada}
\affiliation{Planetary Exploration Research Center (PERC), Chiba Institute of Technology, Tsudanuma 2-17-1, Narashino, Chiba 275-0016, Japan}

\author{Takaya Okamoto}
\affiliation{Planetary Exploration Research Center (PERC), Chiba Institute of Technology, Tsudanuma 2-17-1, Narashino, Chiba 275-0016, Japan}



\begin{abstract}

The asteroid (3200) Phaethon is known to be the parent body of the Geminids, although meteor showers are commonly associated with the activity of periodic comets.
What is most peculiar to the asteroid is its comet-like activity in the ejection of micrometer-sized dust particles at every perihelion passage, while the activity of the asteroid has never been identified outside the near-perihelion zone at $0.14~\mathrm{au}$ from the Sun.
From the theoretical point of view, we argue that the activity of the asteroid is well explained by the electrostatic lofting of micrometer-sized dust particles with the aid of mobile alkali ions at high temperatures.
The mass-loss rates of micrometer-sized particles from the asteroid in our model is entirely consistent with the values inferred from visible observations of Phaethon's dust tail.
For millimeter-sized particles, we predict three orders of magnitudes higher mass-loss rates, which could also account for the total mass of the Geminid meteoroid stream by the electrostatic lofting mechanism.

\end{abstract}



\section{Introduction} \label{sec:intro}

When a comet approaches the Sun, dust particles are ejected from the surface of the comet owing to sublimation of ices, which exerts a drag force on refractory dust particles originally embedded in the ices.
The massive population of the particles, which tend to stay near the orbit of their parent body for an extended period of time, may be observed as meteor showers or fireballs, if the earth intersects their orbits.
In general, parent bodies of meteoroids in meteor showers are periodic comets, but the parent body of the Geminids is exceptionally associated with the asteroid (3200) Phaethon whose perihelion is currently located at $q = 0.140~\mathrm{au}$.
The ejection of micrometer-sized dust particles from Phaethon near its perihelion at heliocentric distances $r_\mathrm{H} = 0.140$--$0.147~\mathrm{au}$ has been recursively observed by the H-1 camera of the SECCHI instrument suite onboard STEREO-A at a wavelength of $0.57~\micron$ \citep{li-jewitt2013,jewitt-et-al2013,hui-li2017}.
As the particles formed a comet-like tail, the ratio $\beta$ of solar radiation pressure to solar gravity is so large ($\beta > 0.07$) that they cannot stay in a bound orbit around the Sun, unlike millimeter-sized particles in the Geminid meteoroid stream.
Since Phaethon appears to be inactive near the Earth orbit ($r_\mathrm{H} \approx 1~\mathrm{au}$), the comet-like activity of the asteroid is confined to the near-perihelion region \citep{jewitt-et-al2018}.
Therefore, there exists some unknown mechanism of dust ejection at work exclusively near $r_\mathrm{H} \approx 0.14~\mathrm{au}$, if Phaethon is an active asteroid rather than a dead comet \citep[see][for the transitional nature of the object]{licandro-et-al2007}.
\citet{jewitt2012} surveyed possible mechanisms of dust ejection from asteroids and noted the importance of cracking and dehydration at high temperatures, but failed to identify the most plausible mechanism of dust ejection \citep[see][for possible progressive production of small grains by fatigue crack growth]{delbo-et-al2014}.
It is worthwhile noting that Phaethon is targeted by JAXA's DESTINY$^+$ mission, scientific objectives of which contain the investigation into the mechanism of dust ejection from the asteroid \citep{arai-et-al2018,arai-et-al2021}.
The unresolved issue of the dust ejection mechanism was recently tackled by \citet{masiero-et-al2021} who postulated that the gradient of vapor pressure caused by sublimation of sodium becomes a driving force to push dust particle outward or inward, depending on the sign of the gradient.
According to their results, meter-sized particles could be lofted off the surface of Phaethon by sodium vapor pressure, but the predominance of such large particles is inconsistent with the long dust tail observed for Phaethon at perihelion.
The observed shape of the dust tail and the dynamics of dust particles ejected from Phaethon imply that the radius of dust particles in the tail lies in the range of a few micrometers \citep[see][]{jewitt-et-al2013}.
Therefore, we must seek the mechanism that lofts micrometer-sized particles over the surface of Phaethon solely in the vicinity of its perihelion to explain the formation of the dust tail.

It should be noted that the electrostatic lofting of dust particles could become a vital mechanism for dust ejection, when certain conditions are simultaneously met \citep{kimura-et-al2014}.
One notable example we could cite is the lunar horizon glow (LHG) that was observed by past lunar missions in the seventies near the terminator of the Moon where the amplification of electric fields is expected \citep{de-criswell1977}.
The LHG is the solar radiation scattered by regolith particles that are lofted off the surface of the Moon near the lunar terminator, while the presence of lofted dust particles around the terminator has also been measured by dust detectors \citep{berg-et-al1976,xie-et-al2020}.
The electrostatic lofting of dust particles takes place under the conditions that water molecules adsorbed on the surface of the particles reduce the cohesive forces between the particles and develop the repulsive forces between the particles owing to the elevation of their electric conductivities \citep{kimura-et-al2014}.
Very recently, a similar mechanism was proposed to operate for the formation of spokes in the B-ring of Saturn where adsorbed oxygen ions assist the ejection of dust particles from the ring near the equinox \citep{hirata-et-al2022}.
By the same token, dust particles could be electrostatically lofted off the surface of Phaethon during its perihelion passage, provided that alkali ions play a role similar to water molecules or oxygen ions.
Indeed, the diffusion of sodium atoms from Na-bearing minerals is accelerated during the perihelion passage and fast enough to reach the surface of dust particles in a micrometer scale \citep{capek-borovicka2009}.
Moreover, the electrical conductivity of feldspar, alkali-ion-bearing silicate, is known to exponentially increase with temperature, because alkali ions in feldspar act as charge carriers at high temperatures \citep{hu-et-al2014}.
Accordingly, we may expect that the diffusion of sodium atoms to the surface of micrometer-sized dust particles greatly reduces the cohesion between the particles and elevates the response of the particles to electric fields in the same way as adsorbed water molecules on the Moon or oxygen ions on the Saturnian B-ring.
In this paper, we examine the hypothesis that the activity of Phaethon near the perihelion is accounted for by the electrostatic lofting of dust particles with the aid of mobile alkali atoms at high temperatures.

\section{Electrostatic lofting of dust particles} \label{sec:model}

\subsection{Forces acting on dust particles}

\subsubsection{Gravitational force} 

We assume that the major constituents of dust particles on the surface of Phaethon are silicate minerals such as feldspar, since less refractory material such as ices and organic matter most likely sublimed away during the past perihelion passages.
This assumption may be justified by the geometric albedo that lies in the range of 0.10--0.16, while the geometric albedo of comets is lower than 0.1, which is typical for carbonaceous matter.
Therefore, we estimate the gravitational force $F_\mathrm{gr}$ acting on dust particles that are situated on the surface of Phaethon on the assumption that the bulk density $\rho$ of the particles is $\rho = 3000~\mathrm{kg~m^{-3}}$ and the mass $M$, the radius $R$, and the porosity $\Pi$ of the asteroid are $M=7.645 \times {10}^{13}~\mathrm{kg}$, $R = 2.3~\mathrm{km}$, and $\Pi = 0.5$, respectively \citep{masiero-et-al2021}.
It should be noted that while the parameters of Phaethon may be still open to debate, the density of the particles is reasonably well constrained by meteor observations \citep{babadzhanov-kokhirova2009}.

\subsubsection{Pressure-gradient force} 

The gradient of sodium vapor pressure induces the pressure-gradient force $F_\mathrm{pg}$ on dust particles where the direction of the force at perihelion is outward at a depth $d \la 50~\mathrm{mm}$ from the surface of Phaethon \citep{masiero-et-al2021}.
According to \citet{masiero-et-al2021}, the outward drag force acting on dust particles per volume $V$ is expected to reach $F_\mathrm{sub}/V \approx 20~\mathrm{N~m^{-3}}$ maximum.
It should be noted the difference between the pressure-gradient forces and the gas-drag forces, because the former is proportional to the volume of dust particles and the latter the cross section of the particles.
We would like to note that the gas-drag forces have been known to be very effective for the ejection of small dust particle, as commonly observed for the formation of cometary comae, while the pressure-gradient forces have never been identified as the major mechanism of dust ejection to date.

\subsubsection{Electrostatic force} 

One may expect that the electrostatic lofting of silicate dust particles is, at first glance, ruled out, because the electric field exerts an attractive force between insulating particles, as opposed to a repulsive force between conductive particles.
Contrary to this expectation, the electrostatic lofting of silicate particles does take effect in the circumstances that ions of charge carriers are present on the surface of the particles \citep{holsteinrathlou-et-al2012}.
Namely, the electrostatic force $F_\mathrm{el}$ on dust particles composed of insulating material turns from attractive to repulsive, when the surface of the particles accumulates charge carriers.
The strength $E$ of electric fields developed near the terminators of the Moon and asteroids found in the literature ranges from $E = 50$--$300~\mathrm{kV~m^{-1}}$ \citep{rennilson-criswell1974,de-criswell1977,lee1996}.
As a conservative estimate, we consider two values of $E = 10$ and $100~\mathrm{kV~m^{-1}}$ for the strength of electric fields that generate electrostatic repulsive forces between dust particles with an accumulation of sodium ions on their surfaces.

\subsubsection{Cohesive force} 

The cohesion is usually the dominant attractive force between micrometer-sized dust particles, while the cohesive force $F_\mathrm{co}$ is proportional to the surface energy.
We assume the surface energy $\gamma_\mathrm{sil}$ of silicate to be $\gamma_\mathrm{sil} = 243.08~\mathrm{mJ~m^{-2}}$, which is equivalent to the surface energy of amorphous silica in a vacuum \citep{kimura-et-al2015,kimura-et-al2020}.
When Na ions migrate to the surface of dust particles, the surface energy  $\gamma$ of the particles may be given by \citep{israelachvili1972}
\begin{eqnarray}
\gamma = \left({\sqrt{\gamma_\mathrm{sil}} - \sqrt{\gamma_\mathrm{Na}}}\right)^2 ,
\end{eqnarray}
where $\gamma_\mathrm{Na}$ is the surface energy of sodium.
We equate the surface energy $\gamma_\mathrm{Na}$ of sodium with the surface tension at absolute zero that is extrapolated from experimental data on surface tension available at various temperatures \citep[e.g.,][]{poindexter-kernaghan1929}.
Because the surface tension $\sigma_\mathrm{Na}$ of sodium is known to vary with the temperature $T$ as \citep{goldman1984}:
\begin{eqnarray}
\sigma_\mathrm{Na} = 240.70~\mathrm{mJ~m^{-2}} \, \left({1 - \frac{T}{2509.5~\mathrm{K}}}\right)^{1.1320} ,
\end{eqnarray}
we obtain $\gamma_\mathrm{Na} = 240.70~\mathrm{mJ~m^{-2}} $ and $\gamma = 5.854 \times {10}^{-3}~\mathrm{mJ~m^{-2}}$.

\subsection{Mass-loss rate}

On the one hand, \citet{jewitt-et-al2013} estimated the mass-loss rate $\dot{M}$ of micrometer-sized dust particles at perihelion from photometric images of the dust tail to be $\dot{M} = 2.89$--$4.63~\mathrm{kg~s^{-1}}$.
On the other hand, the mass-loss rate $\dot{M}$ due to the electrostatic lofting of dust particles near the terminator can be estimated by \citep{kimura-et-al2014}
\begin{eqnarray}
\dot{M} = \frac{4 S}{P \Sigma} \zeta ,
\label{eq:mass-loss-rate}
\end{eqnarray}
where $S$ is the surface area of Phaethon that is swept by the terminator, $P$ is the rotation period of Phaethon, $\Sigma$ is the specific surface area of dust particles, and $\zeta$ is a fraction of $S$ that contributes to the electrostatic lofting of dust particles with an ideal geometry\footnote{An ideal geometry may be achieved with a small rock in tens of millimeters that is located in a dark shadow near the terminator region but in part illuminated by solar radiation \citep{criswell1973,de-criswell1977}.}.
By computing the value of $S$ for a spherical body with the rotation state parameters given by \citet{hanus-et-al2018}, we find the value of $S$ to exceed 99\% of the total surface area of Phaethon near the perihelion.
Therefore, we may simply assume $S=6.648 \times {10}^{7}~\mathrm{m^2}$ and $P = 3.604~\mathrm{hr}$, while we adopt the value of $\Sigma = 0.6 \times {10}^3~\mathrm{m^2~kg^{-1}}$, which is the specific surface area for micrometer-sized particles of the lunar regolith \citep{gammage-holmes1975,masiero-et-al2021}.
Hereafter, the parameter $\zeta$ is treated as a free parameter ($0 \le \zeta \le 1$), since the value of $\zeta$ depends on the unknown topography of Phaethon's surface in tens of millimeter scales and thus an estimate of $\zeta$ is not straightforward.

\section{Results} \label{sec:results}

\begin{figure}
\epsscale{0.5}
\plotone{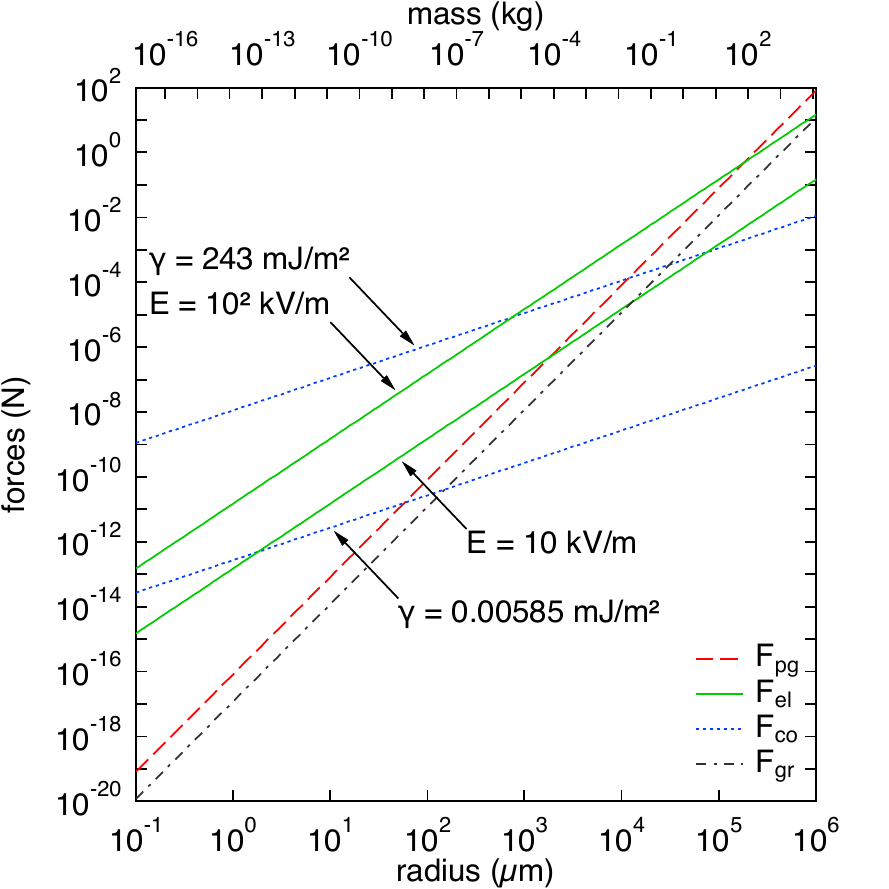}
\caption{Comparison of forces acting on a dust particle as a function of particle radius $r$.
The dashed line: the pressure-gradient force $F_\mathrm{pg}$; the solid line: the electrostatic force $F_\mathrm{el}$; the dotted line: the cohesive force $F_\mathrm{co}$; the dash-dotted line: the gravitational force $F_\mathrm{gr}$.
Two different assumptions on the electric field strength of $E= 10$ and $100~\mathrm{kV~m^{-1}}$ and on the surface energy of $\gamma = 0.00585$ and $243~\mathrm{mJ~m^{-2}}$ are chosen as plausible parameters.
\label{f1}}
\end{figure}
In Fig.~\ref{f1}, we compare the forces acting on a {\sl conductive} dust particle as a function of particle radius at perihelion in the range of radius $r = 0.1$--${10}^{6}~\micron$.
The gravitational ($F_\mathrm{gr}$), pressure-gradient ($F_\mathrm{pg}$), electrostatic ($F_\mathrm{el}$), and cohesive ($F_\mathrm{co}$) forces are given by the dash-dotted, the dashed, the solid, and the dotted lines, respectively.
For the electrostatic and cohesive forces, we plot two lines with different assumptions on the electric field strength of $E= 10$ and $100~\mathrm{kV~m^{-1}}$ and on the surface energy of $\gamma = 0.00585$ and $243~\mathrm{mJ~m^{-2}}$.
The lower surface energy assumes the diffusion of alkali ions to the surface, as a result of which the surface energy is reduced and at the same time the electric conductivity is elevated.
It turns out that the pressure-gradient force is negligible for micrometer-sized dust particles compared to cohesion and electrostatic repulsion, while it exceeds the gravitational force irrespective of particle radius as already noted by \citet{masiero-et-al2021}.
In contrast, electrostatic forces are strong enough to aid the detachment of micrometer-to-millimeter-sized particles from the surface, provided that mobile alkali ions reduce cohesion and elevate the electric conductivity at high temperatures.

\begin{figure}
\epsscale{0.5}
\plotone{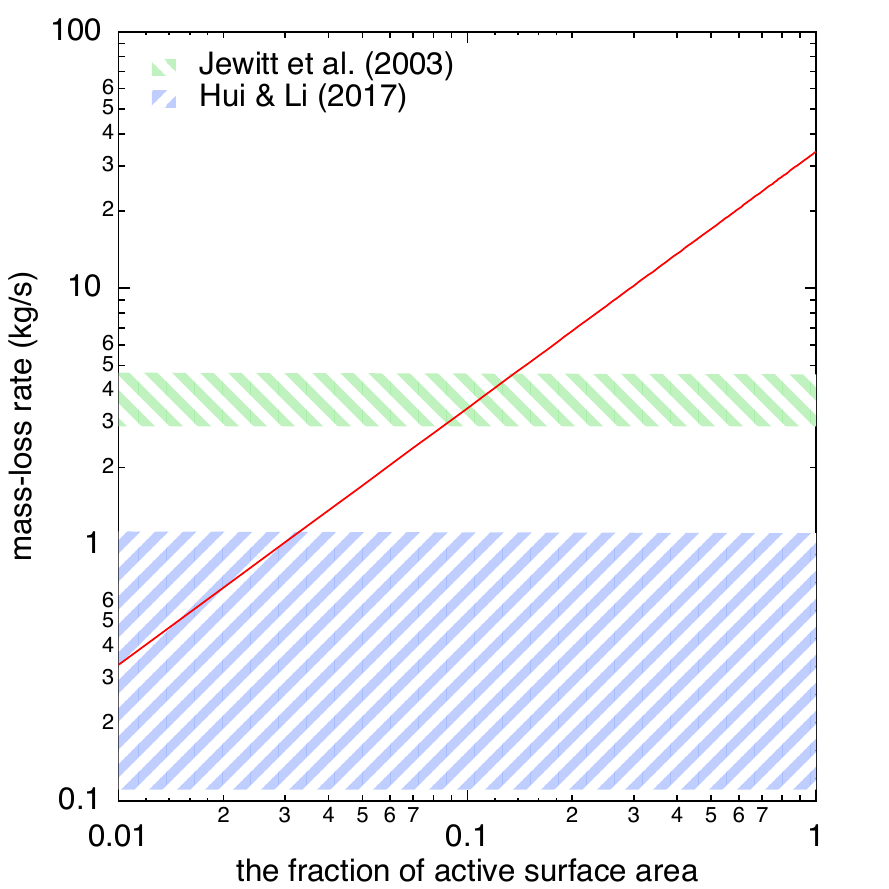}
\caption{The mass-loss rate $\dot{M}$ of micrometer-sized dust particles from the surface of the asteroid (3200) Phaethon (the solid line) as a function of the surface fraction $\zeta$ that contributes to the electrostatic lofting of dust particles with the aid of mobile alkali ions at high temperatures.
The shaded areas indicate the mass-loss rates $\dot{M}$ of micrometer-sized dust particles from the surface of Phaethon at perihelion inferred from photometric images of the Phaethon's dust tail at perihelion \citep{jewitt-et-al2013,hui-li2017}.
\label{f2}}
\end{figure}
Figure~\ref{f2} shows the mass-loss rates $\dot{M}$ (the solid line) predicted by Eq.~(\ref{eq:mass-loss-rate}) as a function of the active surface fraction $\zeta$ that contributes to the electrostatic lofting of dust particles with the aid of mobile alkali ions at high temperatures.
The mass-loss rate $\dot{M}$ of micrometer-sized dust particles from Phaethon lies in the range of $\dot{M} = 0$--$34~\mathrm{kg~s^{-1}}$ when the electrostatic lofting of dust particles is at work.
Also depicted as shaded areas in Fig.~\ref{f2} are the ranges of the mass-loss rates inferred from visible observations of the dust tail that were produced at perihelion passages of Phaethon \citep{jewitt-et-al2013,hui-li2017}.
We find that the observationally-constrained mass-loss rates are consistent with $\zeta \approx 0.01$--$0.1$, implying that most of the surface area fail to develop conditions suitable to launch dust particles by the electrostatic lofting mechanism.

\section{Discussion} \label{sec:discussion}

On the basis of our results, it is feasible that the dust tail of the asteroid Phaethon is produced by electrostatic lofting of micrometer-sized dust particles at perihelion with the aid of mobile alkali ions.
In the case of LEAM experiments on the Moon, \citet{berg-et-al1976} reported a peak of dust ejection around the sunrise,
the full width at half maximum of which is $45.5~\mathrm{hr}$ for the east sensor and $43.3~\mathrm{hr}$ for the up sensor.
By considering the lunar rotation period of $708.7~\mathrm{hr}$, the lower limit $\zeta_{\min}$ to the active surface fraction $\zeta$ may lie in the range of $\zeta_{\min} = 0.061$--$0.064$ for the Moon.
Therefore, the value of $\zeta \approx 0.01$--$0.1$ estimated above is not unrealistic for Phaethon, unless the topography of Phaethon's surface is extremely unfavorable for the electrostatic lofting of dust particles, compared to the lunar surface.
In conclusion, we may state that the electrostatic ejection of dust particles becomes active at approximately $1$--$10\%$ of Phaethon's surface area during the perihelion passage.

Our results indicate that micrometer-to-millimeter-sized particles can be ejected by the electrostatic lofting mechanism with the aid of mobile alkali ions at high temperatures.
It should be, however, noted that the formation of the Geminid meteoroid stream would necessitate higher mass-loss rates, compared to the formation of the dust tail, as suggested by \citet{jewitt-et-al2013}.
On the one hand, the total mass of $M_\mathrm{tot} \approx (0.4$--$1.3) \times {10}^{12}~\mathrm{kg}$ has recently been derived for Phaethon's dust trail from white-light observations by the Wide-field Imager for Parker Solar Probe (WISPR) onboard Parker Solar Probe
\citep{battams-et-al2020}.
Meteor observations also suggest a similar value for the total mass of Geminid meteoroid stream, if the density of Geminids is $\rho = 3000~\mathrm{kg~m^{-3}}$ \citep{ryabova2017}.
On the other hand, the dynamical age $\tau$ of the Geminids would be $\tau \la 600$ or $1000 \la \tau \la 2000~\mathrm{yr}$, depending on the surface-area to mass ratio of the meteoroids \citep{gustafson1989,ryabova1999}.
Therefore, the mass-loss rate of $\dot{M} \approx 3~\mathrm{kg~s^{-1}}$ inferred from observations of Phaethon's dust tail at perihelion amounts only to the total mass of $M_\mathrm{tot} \la 5.7 \times {10}^{10}~\mathrm{kg}$ or $9.5 \times {10}^{10}~\mathrm{kg} \la M_\mathrm{tot} \la 1.9 \times {10}^{11}~\mathrm{kg}$, which is insufficient for the production of the Geminids.
Here, we claim that the mass-loss rates of millimeter-sized particles by electrostatic lofting would be as high as $\dot{M} \approx 3 \times {10}^{3}~\mathrm{kg~s^{-1}}$ at $\zeta \approx 0.1$, since the specific surface area $\Sigma$ of particles is inversely proportional to the size of the particles \citep{carrier-et-al1991}.
If the ejection of millimeter-sized dust particles by the electrostatic lofting mechanism lasts for one day during every perihelion passage over $\tau = 2000~\mathrm{yr}$, we estimate the total mass of $M_\mathrm{tot} \approx 0.4 \times {10}^{12}~\mathrm{kg}$ for the Geminid meteoroid stream, which is within the bounds of the recent total mass estimates \citep[cf.][]{battams-et-al2020}.
Accordingly, the electrostatic lofting of dust particles would offer a feasible explanation for the formation of not only Phaethon's dust tail but also the Geminid meteoroid stream.

Contrary to our model, if the pressure-gradient force were at work as suggested by \citet{masiero-et-al2021}, then decimeter-to-meter-sized particles could be recursively ejected from Phaethon at perihelion.
This significantly increases the risk of a fatal accident in the DESTINY$^+$ mission, since meter-sized particles tend to stay on the orbit of Phaethon and might collide with the spacecraft.
The spacecraft must avoid a collision with a meter-sized particle, because the impact energy corresponds to a kiloton of TNT, implying that the collision results in catastrophic destruction of the spacecraft during a flyby.
In contrast, the electrostatic lofting mechanism at perihelion most likely does not threaten the spacecraft, since the ejection of particles is limited to millimeter or smaller sizes and the probability of an impact with a millimeter-sized particle in Phaethon's dust trail has already been incorporated into the mission plan.
While \citet{ye-et-al2018} provided observational evidence that particles with a radius of $2~\mathrm{m}$ or larger are absent in the Geminid meteoroid stream, \citet{madiedo-et-al2013} reported observational findings of a particle with a radius of $40~\mathrm{mm}$ in the Geminid meteoroid stream, which was observed as a fireball. 
Since these results do not rule out the presence of particles with a radius of $1~\mathrm{m}$, it would be inevitable to conduct a more careful investigation of the dust ejection mechanism on the condition that success in the mission is a must-have.

Phaethon may share the same parent body with asteroids (155140) 2005 UD and (225416) 1999 YC, the former, in particular, is a possible flyby candidate for the extended mission of DESTINY$^+$ \citep{ohtsuka-et-al2006,ohtsuka-et-al2008,arai-et-al2018,arai-et-al2021}.
It has never been proven whether these plausible siblings of Phaethon show a comet-like activity at their perihelia or not, although a lack of dust emissions at aphelia was confirmed observationally \citep{kasuga-jewitt2008}.
However, the electrostatic lofting of dust particles does not seem to currently take place at their perihelia of $q > 0.16~\mathrm{au}$, from which the diffusion of Na is not expected, according to \citet{masiero-et-al2021}.
Nonetheless, 2005 UD is a likely parent body of the Daytime Sextantid meteor shower whose dynamical age of $\tau > 10^{4}~\mathrm{yr}$ was estimated by numerical simulations \citep{jakubik-neslusan2015}.
The dynamical evolution of 2005 UD numerically investigated by \citet{maclennan-et-al2021} reveals that the perihelion of 2005 UD was also located at $q \approx 0.14~\mathrm{au}$ approximately $1.5 \times {10}^{4}~\mathrm{yr}$ ago.
Therefore, we have no reason to rule out the possibility that a dust trail was produced along the trajectory of 2005 UD by the electrostatic lofting mechanism $1.5 \times {10}^{4}~\mathrm{yr}$ ago, which does not contradict the dynamical age of the Daytime Sextantids.
Owing to a lower gravity, the electrostatic lofting mechanism enables to launch larger meteoroids from 2005 UD than Phaethon, in agreement with a model of the Daytime Sextantid meteor shower \citep{jakubik-neslusan2015}.
On the basis of our model, the total mass of the Daytime Sextantids should be straightforwardly estimated, since the orientation of its rotational axis has been derived from photometric observations \citep{huang-et-al2021}.
Although such an estimate is beyond the scope of this paper, a more detailed study on the formation of the Daytime Sextantids would be of great value for the DESTINY$^+$ mission in terms of risk assessment.

\begin{acknowledgments}
H.K. is grateful to the Grants-in-Aid for Scientific Research of the Japan Society for the Promotion of Science (JSPS KAKENHI Grant Number JP21H00050).
\end{acknowledgments}

\end{document}